\documentstyle[12pt,a4]{article}
\newcommand{\bq}{\begin{equation}}
\newcommand{\eq}{\end{equation}}
\newcommand{\ba}{\begin{eqnarray}}
\newcommand{\ea}{\end{eqnarray}}
\newcommand{\mathrm}{\rm}

\newcommand {\zf}{$_{Z\!
F}\!I\!^{\textstyle T}\!\!T\!\!_
{{\textstyle E}\!R}$}

\newcommand{\afb}{A_{FB}}
\newcommand{\oalf}{${\cal O }(\alpha$)$\:$}

\newcommand{\nobody}{\rule{0ex}{1ex}}
\newcommand{\az}{A_0}
\newcommand{\ao}{A_1}

\newcommand{\apol}{A_{pol}}

\newcommand{\swt}{\sin^2 \theta_W}
\newcommand{\rgt}{r_{\gamma}^T}
\newcommand{\rt}{R_T}
\newcommand{\rfb}{R_{FB}}
\newcommand{\rpol}{R_{pol}}
\newcommand{\ra}{R_A}
\newcommand{\rza}{r_0^A}
\newcommand{\rzt}{r_0^T}
\newcommand{\ia}{J_A}

\newcommand{\iT}{J_T}

\newcommand{\st}{\sigma_T}
\newcommand{\sfb}{\sigma_{FB}}
\newcommand{\slr}{\sigma_{LR}}
\newcommand{\spol}{\sigma_{pol}}
\begin{document}
%\begin{titlepage}
%======================================================
  \vspace{-1.4cm}
  \begin{flushright}
  {%\large
 CERN--TH.6590/92 \\
\nobody \\ }
\vspace*{2.0cm}
\end{flushright}
\vfill
\begin{center}
{\Large \bf
CROSS-SECTION ASYMMETRIES
\\
AROUND THE $Z$ PEAK
\vspace*{2cm}\\
}
{\Large \it
    T. Riemann}
\vspace*{1.0cm}\\
%$^1$
Theory Division, CERN, CH-1211 Geneva, Switzerland
\vspace*{.2cm}\\
and
\vspace*{.2cm}\\
DESY --
Institut f\"ur Hochenergiephysik, O-1615 Zeuthen, Germany
\vspace{1.0cm}\\
\date{\today}

\thispagestyle{empty}
%*********************************************
\vfill
{\bf Abstract}
\end{center}
\normalsize
\noindent
A simple model-independent formula for cross-section
asymmetries $A$ in fermion-pair production is derived,
which may be used for the analysis of LEP~1 data, $A = \sum_n
A_n (s - M_Z^2)^n$.
The coefficient $A_0$ depends on the $Z$ boson exchange, $A_1$
additionally on the $\gamma Z$ interference, while the
higher-order contributions are practically redundant.
QED corrections are taken into account, and relations to
other approaches are indicated.

\vspace{2.0cm}
%================================
\vspace{2cm}
\begin{flushleft}
{%\large
CERN--TH.6590/92  \\
%\nobody \\
%\nobody \\ }
 July 1992      \\}
\end{flushleft}
\newpage
\setcounter{page}{1}
%
%-----------------------------------------------------------
%

\section{Introduction}
For fermion-pair production at LEP~1 energies,
%-------------
\bq
e^+ e^- \rightarrow (\gamma, Z) \rightarrow f^+ f^- (\gamma),
\label{e0}
\eq
%------------
experimentalists may present their cross-section asymmetries
in a rather simple way:
%-------------
\bq
A(s) = A_0 + A_1 \left(\frac{s}{M_Z^2} - 1 \right) +
             A_2 \left(\frac{s}{M_Z^2} - 1 \right)^2 + \ldots
\label{e1}
\eq
%------------
Instead of $s$,
one can also use the centre-of-mass energy as a variable:
$(s/M_Z^2-1) = (\sqrt{s}/M_Z+1)(\sqrt{s}/M_Z-1) \sim 2 (\sqrt{s}/M_Z-1)$.
Depending on the accuracy of the data, higher-order terms in the
expansion  may be neglected.
Within $\sqrt{s} = M_Z \pm
 \Gamma_Z$, the $|(s/M_Z^2-1)|$ is less than 0.05, and coefficients
beyond $A_2$ will hardly be within reach.
With the data of the 1990 running period, for the $\tau$-polarization
$A_{pol}$, only the peak value $A_0$ has been determined, and the
forward--backward asymmetry $A_{FB}$ is known to be an almost linear
function of $s$.
The LEP~1 data of 1992 will be much more precise and it seems to be
reasonable to analyse the shapes of the cross-section asymmetries,
and not only the peak values.

In this letter, it will be proved that~(\ref{e1}) is a unique,
model-independent
ansatz for cross-section asymmetries around the $Z$ peak.
The coefficients $A_n$ contain the complete physical information without
approximations.
Allowing for a smooth (and calculable) dependence of the $A_n$ on $s$,
the complete QED corrections may be included in the ansatz~(\ref{e1}).
We derive~(\ref{e1}) in two steps, first neglecting the photonic
corrections.

\section{Model-independent approach to asymmetries}
A cross section may be parametrized in a unique way:
%-------------
\bq
\sigma (s) = \frac{4}{3} \pi \alpha^2
\left[ \frac{r_{\gamma}}{s} +
\frac {s R + (s - M_Z^2) J} {(s-M_Z^2)^2 + M_Z^2 \Gamma_Z^2}
+
\sum_n \frac{r_n}{M_Z^2} \left( \frac{s}{M_Z^2} - 1 \right)^n \right].
\label{e3}
\eq
%------------
For the total cross section $\sigma_T$, a formula like~(\ref{e3})
has been extracted
from the Standard Model in~\cite{borrelli}.
In~\cite{smatrix}, this formula has been derived
for $\sigma_T$, but also for left--right ($\sigma_{pol}, \sigma_{LR}$)
and forward--backward ($\sigma_{FB}$) cross-section differences
from an ansatz for the scattering matrix.
The
definitions of mass and width are also discussed there.
At LEP~1, the corresponding asymmetries are:
%-------------
\bq
\afb = \frac{\sigma_{FB}}{\sigma_T},
\hspace{1.cm}
\apol = \frac{\sigma_{pol}}{\sigma_T}.
\label{e4}
\eq
%------------
The four independent matrix elements ${\cal M}_i$ for the scattering
of helicity states,
%---
\bq
{\cal M}_i (s) = \frac{R_{\gamma}}{s} + \frac{R_Z^i}{s-s_Z}
+ \sum_n F_i^{(n)}(s_Z)(s-s_Z)^n,
\label{e2b}
\eq
%--
with  $s_Z=M_Z^2-iM_Z\Gamma_Z$, are used to derive the cross sections.
The $R_{\gamma}, R_Z^i,$ and $F_i^{(n)}$ are complex constants,
which characterize the scattering process. They compose the parameters
$r_{\gamma}, R, J, r_n$ in~(\ref{e3}) in $\sigma_T, \sigma_{pol},
\sigma_{FB}$~\cite{smatrix}.
For definiteness, we quote the quantum-mechanical interpretation of the
coefficients:
%-------------
\bq
r_{\gamma}^T = \left| \frac{\alpha(s)}{\alpha} \right|^2  Q_e^2 Q_f^2
             \sim 1.131 Q_F^2,
\hspace{1.cm}
r_{\gamma}^{FB} = r_{\gamma}^{pol} = 0,
\label{e5}
\eq
%------------
\bq
R_T = \kappa^2 (a_e^2+v_e^2)(a_f^2+v_f^2)
+ 2 \kappa |Q_e Q_f| v_e v_f \frac{\Gamma_Z}{M_Z}
\Im m \frac{\alpha(s)}{\alpha},
\nonumber
\label{e6a}
\eq
%------------
\bq
R_{FB} = 3 \kappa^2 a_e v_e a_f v_f
+ \frac{3}{2} \kappa |Q_e Q_f| a_e a_f \frac{\Gamma_Z}{M_Z}
\Im m \frac{\alpha(s)}{\alpha},
\nonumber
\label{e6b}
\eq
%------------
\bq
R_{pol} = - 2 \kappa^2 (a_e^2+v_e^2) a_{\tau} v_{\tau}
- 2 \kappa |Q_e Q_f| v_e a_{\tau} \frac{\Gamma_Z}{M_Z}
\Im m \frac{\alpha(s)}{\alpha},
\label{e6}
\eq
%------------
\bq
J_A  = 2 |Q_e Q_f| \Re e \frac{\alpha(s)}{\alpha} \kappa \kappa_A,
\nonumber
\label{e8a}
\eq
%------------
\bq
\kappa_T      = v_e v_f,  \hspace{1.cm}
\kappa_{FB}   = \frac{3}{4} a_e a_f,  \hspace{1.cm}
\kappa_{pol} = - v_e a_{\tau},
\label{e8}
\eq
%------------
\bq
\kappa = \frac{G_{\mu}}{\sqrt{2}} \frac{M_Z^2}{8 \pi \alpha}
       = 0.3724 \left( \frac{M_Z}{91} \right)^2.
\label{e9}
\eq
%------------
The parameter $R$ in~(\ref{e3}) is built    out of the residua of the
$Z$ pole in the four              helicity amplitudes, %for details
the $J$ comes from the $\gamma Z$ interference, and $\alpha(s)$ is the
running QED coupling constant, which we assume to be known.
\\
Explicit expressions for the coefficients $R$ and $J$ in
$\st, \sfb, \slr$ and $\spol$ in terms
of the Standard Model have been derived in~\cite{stukni}.
The $r_n$ are additional, non-resonating quantum corrections and yield
negligible contributions.

{}From~(\ref{e3}) and~(\ref{e4}), one easily derives expressions for
the coefficients $A_n$ in~(\ref{e1}).
The peak value of the asymmetry is (index $A=FB,pol$):
%------------
\bq
A_0 = \frac{ \ra + \gamma^2 \rza} { \rt + \gamma^2 (\rzt + \rgt)}
\sim
\frac{\ra}{\rt + \gamma^2 \rgt} \sim \frac{\ra}{\rt},
\label{e15}
\eq
%------------
where it is taken into account that $r_{\gamma}^A=0$.
Then,     non-resonating quantum corrections
$\gamma^2 r_0 \sim \frac{\alpha}{\pi} \gamma^2 \sim 2 \times 10^{-6}$
are neglected,
and finally also the pure photonic contribution $\rgt$,
                               which is multiplied with $\gamma^2
%------------
= \Gamma_Z^2 / M_Z^2 \sim 0.75 \times 10^{-3}$.
For the coefficient $\ao$, again after safely neglecting the quantum
corrections $r_0, r_1$, one gets:
%------------
\ba
\ao =
\left[ \frac{J_A}{R_A} -  \frac{J_T}{R_T+\gamma^2 \rgt}
       + \frac{2 \gamma^2 \rgt}{\rt + \gamma^2 \rgt}
\right] \az % 2 \gamma^2
\sim
\left[ \frac{J_A}{R_A} -  \frac{J_T}{R_T} \right] \az.
\label{e16}
\ea
%------------
The higher-order coefficients are defined by a recurrence relation.
                   Neglecting again the $r_n$ and terms of
order ${\cal O}(\gamma^2)$,
%------
\bq
A_n =
- \left( 2+\frac{J_T}{R_T} \right) A_{n-1}
+ \left[ \left( 1+\frac{J_A}{R_A} \right) \delta_{n,2}
   - \left( 1 + \frac{J_T + r_{\gamma}^T}{R_T} \right) \right] A_{n-2}.
\label{23a}
\eq
%-------

Assuming for a moment that photonic corrections may be neglected,
and that the asymmetry~(\ref{e1}) may be interpreted in terms of vector-
and axial-vector couplings, it is:
%-------------
\bq
A_0^{FB} = 3 \frac{a_e v_e a_f v_f}{(a_e^2+v_e^2)(a_f^2+v_f^2)},
\hspace{1.cm}
A_0^{pol} = - \frac{2 a_{\tau} v_{\tau}}  {a_{\tau}^2+v_{\tau}^2}.
\label{e2}
\eq
%------------
A notation is used with $a_f=1, v_f=1-4|Q_f|\swt$. Further,
%------------
\ba
A_1^{FB} \sim \frac{3}{2 \kappa} |Q_e Q_f| a_e a_f
\frac{  (a_e^2+v_e^2)(a_f^2+v_f^2) -  4 v_e^2 v_f^2}
     {(a_e^2+v_e^2)^2 (a_f^2+v_f^2)^2             },
\label{e16a}
\ea
%------------
\ba
A_1^{pol} \sim - \frac{2}{\kappa} |Q_e Q_{\tau}| \frac{v_e a_{\tau}
       (a_{\tau}^2-v_{\tau}^2) }
     {(a_e^2+v_e^2)   (a_{\tau}^2+v_{\tau}^2)^2 }.
\label{e16b}
\ea
%------------
The $A_0$ is completely determined by the residua of the $Z$ resonance.
It is not influenced by the $\gamma Z$ interferences, while the $A_1$
gets its leading contributions just from them.
With a rising accuracy of the data, the photonic contributions
may not be neglected in the above definitions~\cite{sommi}.
 In $\ao$, there is an additional suppression due to
   the small factor $(s/M_Z^2-1)$.
With ideal data, the interpretation of the asymmetry is straightforward.
The $M_Z, \Gamma_Z, \rt$ and $\iT$ may be determined
 from an analysis of the $Z$ line shape\footnote{
A serious line-shape analysis has at least four free parameters.
While $M_Z$ and $\Gamma_Z$ are universal,
the $\rt, \iT$ are different for the
different channels.}.
Then, $A_0$ allows the determination of
$\ra$, and afterwards            the
$\gamma Z$ interference $\ia$ may be derived from $A_1$.
Instead of $R_{A}$, we may determine the ratio $R_A/R_T$,
which allows us to determine other coupling combinations than may
be obtained from a line-shape anlalysis, thus improving determinations
of the effective weak mixing angle. The problems connected with a
secondary interpretation of the model-independent findings will not
be discussed here, although they are important and interesting.

Since the
$A_0$ is independent of the $\gamma Z$ interferences, it is stable
against several phenomena.
For the
$Z$~peak position $s_p$, one may derive the relation~\cite{jegerl}:
%------------
\bq
\Delta \sqrt{s_p} = \Delta M_Z + \frac{1}{4} \gamma^2 {M_Z} \, \,
\Delta    \left( \frac{\iT}{\rt} \right) + \ldots
\label{e17}
\eq
%------------
between an                              uncertainty in $M_Z$ and  an
uncertainty in the $\gamma Z$ interference.
The latter influences $A_1$.
Similarly, for a hypothetical heavy gauge boson $Z'$, the effects from
its virtual exchange transform after a partial fraction decomposition
into simple shifts of the $\gamma Z$
interferences~\cite{jegerl}:
%------------
\bq
\Delta \left( \frac{\iT}{\rt} \right) =
- 2 \frac{g'^2}{g^2} \frac{M_{Z'}^2}{M_{Z'}^2-M_{Z}^2}
\frac {(a_e a_e' + v_e v_e')(a_f a_f' + v_f v_f')}
      {(a_e^2    + v_e^2   )(a_f^2    + v_f^2   )} ,
\label{e18}
\eq
%------------
and analogously for the cross-section differences.
Again,
the $A_1$ will be influenced, while $A_0$ is sensitive exclusively to
the $Z Z'$ mixing effect.
The $\rgt$ in the definition of $A_2$
is of the order of one, thus not       suppressed by
neglecting    $\gamma^2$.
As mentioned above, the whole $A_2$ is suppressed at LEP~1 with a factor
of $(s/M_Z^2-1)^2 \leq 0.3\%$ and it contains no new physical information
with respect  to $A_{0,1}$.

\section{Inclusion of photonic corrections}
In this section, the modifications             from the photonic
corrections are discussed.
Neglecting the initial--final interference bremsstrahlung\footnote{
These very small corrections may be properly
taken into account~\cite{smatrix}.
Then, the number of free parameters increases.},
the cross sections $\sigma$ [eq.~(\ref{e3})] may be replaced by
%------------
\bq
{\bar{\sigma}} (s) = \int dk \sigma (s') \rho (k),
\label{e19}
\eq
%------------
where $s'=(1-k)s$, and the radiator function $\rho (k)$ contains the
QED corrections.
An         introduction to photonic corrections in the
language of scattering
amplitudes is given in~\cite{cargese}.
A rather complete discussion of~(\ref{e19}) may be found
in~\cite{yr,npb351},
and in the references therein.
For the present purpose, after inserting~(\ref{e3}) into~(\ref{e19}),
the cross section may be rewritten as follows:
%------------
\bq
{\bar{\sigma}} (s)
 = \frac{4}{3} \pi \alpha^2
\left[ \frac{ {\bar r}_{\gamma}}{s} +
\frac {s {\bar R} + (s - M_Z^2) {\bar J}} {(s-M_Z^2)^2 + M_Z^2 \Gamma_Z^2}
+
\sum_n
\frac{{\bar r}_n}{M_Z^2} \left( \frac{s}{M_Z^2} - 1 \right)^n \right].
\label{e20}
\eq
%------------
Any of the barred
parameters differs from its unbarred partner by a correction
factor $C(s)$:
%------------
\bq
{\bar{R}} = C_R(s) \, R, \hspace{1.cm}
C_R(s) = {\cal I}
\left[ \frac{s'}{s} \, \,
\frac{(s -M_Z^2)^2+M_Z^2 \Gamma_Z^2}
                              {(s'-M_Z^2)^2+M_Z^2 \Gamma_Z^2} \right],
\label{e21}
\eq
%------------
\bq
{\cal I} [B] = \int dk B(s') \rho (k).
\label{e22}
\eq
%------------
The other correction factors are analogously defined:
%------------
\ba
C_J(s)&=&{\cal I} \left[
      \frac{s'-M_Z^2}{s-M_Z^2}
\, \,
\frac{(s -M_Z^2)^2+M_Z^2 \Gamma_Z^2}
                              {(s'-M_Z^2)^2+M_Z^2 \Gamma_Z^2}
\right] ,
\\
C_r(s)&=&{\cal I} \left[ \frac{s}{s'} \right], %hspace{1.cm}
\\
C_n(s)&=&{\cal I} \left[ \frac{(s'-M_Z^2)^n}{(s-M_Z^2)^n} \right].
\label{e23}
\ea
%------------
The reader may wonder that some of the corrections seem to be singular at
$\sqrt{s}=M_Z$.
This is not the case for the products $C_{J,n}(s) (s-M_Z^2)$.
As may be seen from the corresponding definitions,
these remain small compared with  e.g. $A_0$, when
$\sqrt{s}$ approaches $M_Z$. At this energy,
the asymmetry is defined
             as the (smooth) limit from the neighbouring energies.
In a more elegant notation, but unnecessarily sophisticated for
applications,
          one could rewrite the asymmetry as a series in powers
of $(s-s_Z)$ as~(\ref{e2b}),           thus regularizing the $C_{J,n}$.
We remark that after inclusion of
photonic corrections the parameters $J$ and $r_n$ may yield non-vanishing
contributions at
$s=M_Z^2$.
This was not the case without the QED corrections, which smear out the
effective energy.

The   QED corrections are well-defined as soon as $M_Z$ and $\Gamma_Z$
are known.
They are
independent of the underlying dynamics of the scattering process.
It is not
difficult to collect explicit expressions for the $C$ functions
from the literature.
They, of course, depend on the handling of the
photonic phase spase,
the inclusion of higher orders, and on acceptance cuts.
For example, the initial-state corrections
in $A_0$ (with possible inclusion of soft photon
exponentiation), with a cut on the energy of the emitted photon, are
$C_R^T = R_T^e(1,1), C_R^{FB} = R_{FB}^e(1,1)$,
where the $R_{T, FB}$ are defined in eqs.~(56)--(57) and~(78)
of~\cite{npb351}.
In the simplest case [initial state radiation to \oalf without cuts],
it is $C_R^T=1+\frac{\alpha}{\pi}H_0^T$,
$C_{R}^{FB}=1+\frac{4}{3}\frac{\alpha}{\pi}H_3^T$, with the $H_{0,3}$
to be taken
from~\cite{663}.

Taking into account the QED corrections,
the experimental data may be fitted with the ansatz
%-------------
\bq
{\bar A}(s) = {\bar A}_0 + {\bar A}_1 \left(\frac{s}{M_Z^2} - 1 \right) +
             {\bar A}_2 \left(\frac{s}{M_Z^2} - 1 \right)^2 + \ldots
\label{e24}
\eq
%------------
The ${\bar A}_n$
may be obtained from the $A_n$, replacing everywhere in the
corresponding definitions the unbarrred variables by barred ones.
For the leading contribution to the forward--backward asymmetry, the
explicit expression is:
%-------------
\bq
{\bar A}_0^{FB} = \frac{C_R^{FB}(s)}{C_R^{T}(s)}
\frac{\rfb}{\rt + [C_r^{T}(s) / C_R^{T}(s)] \gamma^2 \rgt}
\sim
%frac{C_R^{FB}(s)}{C_R^{T}(s)}
0.998 \frac{\rfb}{\rt + 0.001}.
\label{e25}
\eq
%------------
Note that the radiator function $\rho_{FB}(k)$ in (\ref{e22}),
which must be used  for the calculation of
$\sigma_{FB}$, differs from the radiator $\rho_T(k)$.
The latter
is used both for $\sigma_T$ and $\sigma_{pol}$, and the expression
for $\apol$ simplifies correspondingly.
The leading term is:
%-------------
\bq
{\bar A}_0^{pol} =
\frac{\rpol}{\rt + [C_r^{T}(s) / C_R^{T}(s)] \gamma^2 \rgt}
\sim  \frac{\rpol}{\rt + 0.001}.
\label{e27}
\eq
%------------
Further, neglecting the strongly suppressed contributions to $A_1$
(index $A=FB, pol$):
%------------
\ba
{\bar{\ao}}=
 \frac{C_J^T(s)}{C_R^T(s)}
\left[ \frac{J_A}{R_A} -  \frac{J_T}{R_T}
\right] {\bar{\az}}.
\label{e31a}
\ea
%------------

The explicit numerical values in~(\ref{e25}) and~(\ref{e27}) may be
taken from Table~1, which is  calculated with
                              the FORTRAN
package \zf~\cite{npb351,plb255,zfitter}.

%\renewcommand{\baselinestretch}{1.0}
%--------------------------
\footnotesize
\begin{table}[thbp]\centering
\begin{tabular}{|c|c|c|c|c|c|c|c|}
\hline
 & & & & & & &  \\
%thickline
%vspace*{1.0mm}
$\sqrt{s}$
%\cline{2-7}
           & $M_Z-2\Gamma_Z$
           & $M_Z-\Gamma_Z$ & $M_Z-\frac{1}{2}\Gamma_Z$ & $M_Z$
           & $M_Z+\frac{1}{2}\Gamma_Z$
           & $M_Z+           \Gamma_Z$
           & $M_Z+2          \Gamma_Z$                   \\
 & & & & & & &  \\
\hline
 & & & & & & &  \\
$C_R^{FB}$
&0.7784 &0.7331  &0.7078  & 0.7350 & 0.9367 &1.2209& 1.8120\\
 & & & & & & & \\
$C_R^{FB}/C_R^{T}$
&0.9977 & 0.9980 & 0.9982 & 0.9982 & 0.9981 &0.9978 &0.9964\\
 & & & & & & & \\
\hline
 & & & & & & & \\
$              C_{\gamma}^T/C_R^T$
&1.7422      & 1.8565 & 1.9264 & 1.8582 & 1.4601 &1.1215 &1.7569 \\
 & & & & & & & \\
\hline
 & & & & & & & \\
$C_J^{FB}/C_R^{FB}$
&1.0881   &1.1111&1.1592& $[-0.004]$
%-4.8113
& 0.6176 &0.5649 &0.4384 \\
 & & & & & & & \\
$C_J^T/C_R^T$
&1.0905  &1.1130    &1.1610    & $ [-0.004]$ & 0.6159 & 0.5631&0.4358\\
 & & & & & & & \\
\hline
\end{tabular}
\caption[ ]
{\it
QED corrections to the parameters of the model-independent
asymmetry formulae; $M_Z = 91.146$ GeV, $\Gamma_Z = 2.499$ GeV.
}
\label{ta6}
\end{table}
%renewcommand{\baselinestretch}{1.2}
\normalsize
%-----------------------

For this purpose, one should use the branch which relies on the
S-matrix ansatz~\cite{zfafb}.
                                       The maximal acollinearity
of the final-state fermions is assumed to be $\xi=15^{\circ}$, and
the minimal energy of one of the fermions to be $E_{\min}= 20 $ GeV
(standard cuts of \zf\ with flag {\tt ZUCUTS}=1).
The photonic corrections to the asymmetries are remarkably stable
against a variation of the cut conditions.
              Higher-order corrections with common exponentiation
of initial- and final-state corrections are taken into account.

The photonic corrections to $A_0$ are nearly negligible.
The reason is that the corrections $C^T$ and $C^{FB}$
differ only owing to hard-photon emission, which is strongly suppressed
at the $Z$ peak~\cite{plb229}.
Explicit expressions for their ratio may be found in eqs.~(81)--(86)
of~\cite{npb351}.
The photonic corrections to $A_1$ show a completely different behaviour.
This is due to the ratio $C_J(s)/C_R(s)$.
In~(\ref{e31a}) we took into account that this ratio
       is nearly identical
for $\sigma_{FB}$ and $\sigma_{T, pol}$.
As has been discussed in~\cite{npb351,zfpc51}, there is an essential
difference between the two corrections $C_J(s)$ and $C_R(s)$:
while the pure $Z$ exchange
cross section (i.e. $C_R$) develops a radiative tail, the $\gamma Z$
interference (i.e. $C_J$) does not.
Consequently, their ratio is
smooth and of order one below the resonance, while above it becomes
considerably smaller since $C_R(s)$ grows up.
Mainly for this reason, the measured asymmetries
are nearly linear functions of $\sqrt{s}$ at $\sqrt{s} < M_Z$, and
become suppressed beyond the peak.
In principle,
the radiative tail may be avoided by a cut on the allowed energy of
the emitted photons~\cite{zfpc51}:
%------------
\bq
\frac{E_{\gamma}}{E_{\mathrm{beam}}} < \Delta = 1 - \frac{M_Z^2}{S}.
\label{e32a}
\eq
%--------------------
At LEP~1, where $s$ is near to $M_Z^2$, this condition is rather
restrictive; e.g. at $\sqrt{s} = M_Z + 2 \Gamma_Z$, it is
$\Delta = 0.1$. Thus, usually one presents data including radiative
corrections (see e.g. figures~19a-c of~\cite{zfpc252}.
In the present approach, the ratio $C_J/C_R$ is the only QED
correction, which is essentially energy dependent. As has been mentioned
above, near $\sqrt{s} = M_Z$, one should better enumerate the smooth
product $(s/M_Z^2-1)C_J/C_R$. This has been done in Table~1; see the
numbers in square brackets there.

The higher-order coefficients $A_n$ are composed out of the
two first ones, and the same is true for their photonic    corrections.

\section{Discussion}
{}From the model-independent $Z$ line-shape formulae, we derived
the corresponding expressions for the forward--backward asymmetry $\afb$
and the $\tau$  polarization $\apol$ at LEP~1 energies.
The analytic expressions are valid for the leptonic and $b$-quark
forward--backward
asymmetries as well as for the $\tau$ polarization.
The remarkably simple power series in $(s-M_Z^2)$
may cover in their coefficients the photonic
corrections as complete as the line-shape formulae do.
The asymmetries are defined by only two free parameters
                                             $A_0$ and $A_1/A_0$
or, alternatively,
$R_A, J_A$ ($A=FB, pol$). The latter ones
      are related to the               $ZZ$ and $\gamma Z$ contributions
to the corresponding asymmetric cross-section combinations.
In a subsequent step, one may determine effective couplings  or
radiatively corrected Standard Model parameters from the
measured model-independent numbers.
Additional, non-resonating quantum corrections may be neglected.
The                   photonic corrections are
defined such that they depend exclusively on $s, M_Z, \Gamma_Z$.
             The FORTRAN package \zf\ may be used for their calculation.
For $A_0$ they are extremely small. The corrections to $A_1$
are dominated by the radiative tail.

{}From a {\em combined} analysis of the line shape and of asymmetries,
one may try to determine the basic quantities of the S-matrix approach,
i.e. the four complex residua of
the $Z$-boson pole $R_Z^i$ in~(\ref{e2b}).
For a given channel, this deserves the measurement of four
independent sets of parameters $R, J$.
              For lepton-pair production, the three energy-dependent
quantities $\sigma_T, A_{FB}, A_{pol}$ have been determined
experimentally. These are sufficient for at least the determination
of the real parts of the leptonic $R_Z^i$.
As long as there is no beam polarization available at LEP~1,
one could try to measure as a fourth, independent leptonic
observable the $\tau$ polarization in forward direction,
$\lambda_{\tau}^F$, or the asymmetry
$\lambda_{\tau}^{FB}$, as was proposed in~\cite{zfitter,rstau}.

\section*{Acknowledgements}
We have been motivated to work out the details of the present approach to
asymmetries by discussions with G.~Altarelli, L.~Maiani, S.~Kirsch and
M.~Pohl. Thanks to A.~Leike and S.~Kirsch for discussions and a careful
reading of the manuscript.

%------------------------
\end{document}